\documentstyle[12pt]{article}
\begin{document}
\begin{center}
\LARGE
{\bf Effect of Boundary Conditions on Cellular Automata that Classify Density}\\

{\bf N. Sukumar}\\

\large
Alexander von Humboldt Fellow\\
Institut f\"{u}r physikalische und theoretische Chemie\\
Universit\"{a}t Bonn\\
Wegelerstra{\ss}e 12, D-53115 Bonn, Germany\\
Permanent address : Theorie International\\
175/1 Lloyds Road, Gopalapuram, Chennai 600086 India
\newpage
{\bf ABSTRACT}\\
\end{center}
\Large

The properties of two-state nearest-neighbour cellular automata (CA) that are capable of
density classification are discussed. It is shown that these CA actually conserve the
total density, rather than merely classifying it. This is also the criterion for any CA
simulation of DFT. The effect of boundary and periodicity
conditions upon the evolution of such CA are elaborated by considering linear and cyclic
lattices and boundary conditions consistent with the conservation criterion. In a bounded
linear lattice, it is possible to achieve a configuration with a single 01 and no 10 domain
wall, or {\it vice versa}, but this is not possible in a cyclic lattice, where these two
domain walls have to appear in pairs. This determines the final stable state of the
automaton.
\newpage

		It has recently been shown by Capcarrere, Sipper and Tomassini$^2$ that
certain two-state r=1 Cellular Automata (CA), namely rules 184 and 226 in
Wolfram's classification$^3$, can perfectly classify the density, in contrast to a result
proved earlier by Land and Belew$^4$. This is due to a redefinition of the
classification criterion for the final configuration in ref.2. The final configurations for
these CA consist of a background of alternating bits, with a block of one or more
adjacent 0's or 1's, depending upon whether the density of 1's in the initial
configuration was less than or greater than 0.5. No other CA performs this density
classification with comparable accuracy: rules 57 and 99 can classify the density
with only 60\% accuracy.\\

		In this paper I point out the properties of the CA that make such density
classification possible and the effect of boundary conditions upon the evolution of
such CA. First consider the same periodic boundary conditions ($s_{N+1}=s_i$ where $s_i$
is the value of site i) employed in ref.2 and define a "vacuum state" consisting
purely of alternating 0's and 1's. The density classification is performed by the
mutual annihilation of quasiparticles (two or more adjacent 1's) and holes (two or
more adjacent 0's), which move in opposite directions through the lattice, until
there are no further quasiparticles or no further holes, depending upon which were
fewer in the initial configuration. A pair of adjacent 1's and a pair of adjacent 0's
(in a background of alternating bits) are topological solitons : one is a kink and the
other an antikink. The topology is identical to that of quasi-1-D polymers like
polyacetylene, where solitons have been identified and simulated by CA$^5$.\\

		During the evolution of the CA rules 184 or 226, the number of 0's and 1's
are each conserved (so these CA are not merely classifying the density, but
rather conserving the density). Such conservative CA were first discussed by
Vichniac$^6$. This conservative evolution is possible because CA rules 184 and 226
can be reformulated as two-site density interchanges or two-site spin flips (if the
bits are visulaized as spins). Rule 184 may be restated :
\begin{equation}
{\rm Flip\ any\ 10 --->  01 ;\ All\ else\ unchanged}
\end{equation}
while rule 226 may be restated as :
\begin{equation}
{\rm Flip\ any\ 01 --->  10 ;\ All\ else\ unchanged}
\end{equation}
This makes 1's and 0's (or opposite spins) propagate in opposite directions, unlike
other conservative CA (rules 170 or 240) where both species drift in the same
direction. Note that rules 57 and 99 are not perfect density classifiers, since they
donot conserve the density (they change a block of contiguous 0's into a block of
1's and vice versa, at each step).\\

		The final state $\Phi_{alt}$ of a CA (rule 184 or 226) consisting of N sites, with
average density $\rho$ = 0.5, is achieved within N/2 steps and oscillates between a
configuration where $s_i=0$ for all even i, $s_i=1$ for all odd i and a configuration where
$s_i=1$ for all even i, $s_i=0$ for all odd i (an "antiferromagnetic" state in the spin
analogy). When the density is different from 0.5 there are solitonic structures,
consisting of either only 1's (for $\rho > 0.5$) or only 0's (for $\rho < 0.5$) superposed upon
$\Phi_{alt}$ , which propagate continuously around the lattice.\\

		The situation changes markedly when the boundary condition is no longer
periodic. Another set of possible boundary conditions consistent with the criterion
of density conservation is :
\begin{equation}
s_0 = 0,\ s_{N+1} = 1\ {\rm for\ Rule\ 184}
\end{equation}
\begin{equation}
s_0 = 1,\ s_{N+1} = 0\ {\rm for\ Rule\ 226}
\end{equation}
With a suitably bounded linear lattice, the "antiferromagnetic" state $\Phi_{alt}$ (obtained
with the periodic lattice) is no longer stable. $\Phi_{alt}$ evolves further, to give a state
consisting of a single block ($\Phi_o$) of 0's and a single block ($\Phi_1$) of 1s
("ferromagnetic" domains), which is then a stable end point of the CA evolution.
The entire evolution is completed within N-1 steps. Writing $W_r^T|\phi(m,n)>$ for the
application of Wolfram's CA rule r upon an arbitrary initial configuration $|\phi(m,n)>$
consisting of m 0's and n 1's,  for T steps, we now have
\begin{equation}
W_{184}^T|\phi(m,n)> = |\Phi_o(m)\Phi_1(n)>\ {\rm for\ all\ } T > N-1	
\end{equation}
where the block of m 0's appears to the left of the block of n 1's, and
\begin{equation}
W_{226}^T|\phi(m,n)> = |\Phi_1(n)\Phi_o(m)>\ {\rm for\ all\ } T > N-1	
\end{equation}
where the block of m 0's appears to the right of the block of n 1's. Thus $W_{184}^{N-1}$
and $W_{226}^{N-1}$ act as ordering operators (as in the normal ordering of creation and
annihilation operators in many body theory).\\

		The effect of the boundary conditions upon the final configurations of the
CA arises from the changed topology of the lattice. From equation (1), it is seen
that the 10 domain wall is unstable under $W_{184}$ while the 01 domain wall is stable.
In a bounded linear lattice, it is possible to achieve an arrangement with a single
01 and no 10 domain wall (as in the RHS of equation 5), but this is not possible in
a periodic lattice (which has the topology of a circle) where these two domain
walls have to appear in pairs. Likewise the 10 domain wall is stable under $W_{226}$
while the 01 domain wall is unstable. As a result $\Phi_{alt}$ is unstable under rules
184 and 226 in a linear lattice with the boundary conditions (3) or (4).

\newpage
\large
{\bf REFERENCES} :\\
\begin{enumerate}
\item H. Singh, N. Sukumar and B. M. Deb, "{\it Atom as a Complex System : One- and
Two-Dimensional Cellular Automata Simulations}" Int. J. Quantum Chem. {\bf 60}, 21 (1996)
\item M. S. Capcarrere, M. Sipper and M. Tomassini, Phys. Rev. Lett. {\bf 77}, 4969 (1996)
\item S. Wolfram, Physica {\bf 10 D}, 1 (1984); S. Wolfram, ed. "{\it Theory and
Applications of Cellular Automata}" (World Scientific, Singapore, 1986)
\item M. Land and R. K. Belew, Phys. Rev. Lett. {\bf 74}, 5148 (1995)
\item N. Sukumar, in "{\it New Challenges in Computational Quantum Chemistry}",
ed. by P. Bagus and R. Broer (University of Groningen, The Netherlands, 1994)
and references therein.
\item G. Y. Vichniac, Physica {\bf 10 D}, 96 (1984)
\end{enumerate}
\end{document}